\PassOptionsToPackage{svgnames,dvipsnames}{xcolor}
\documentclass[twocolumn]{aastex7}
\usepackage{CJK}

\definecolor{linkcolor}{cmyk}{1,1,0,0}
\definecolor{mplaliceblue}{rgb}{0.941176470588,0.972549019608,1.0}
\definecolor{mplantiquewhite}{rgb}{0.980392156863,0.921568627451,0.843137254902}
\definecolor{mplaqua}{rgb}{0.0,1.0,1.0}
\definecolor{mplaquamarine}{rgb}{0.498039215686,1.0,0.83137254902}
\definecolor{mplazure}{rgb}{0.941176470588,1.0,1.0}
\definecolor{mplbeige}{rgb}{0.960784313725,0.960784313725,0.862745098039}
\definecolor{mplbisque}{rgb}{1.0,0.894117647059,0.76862745098}
\definecolor{mplblack}{rgb}{0.0,0.0,0.0}
\definecolor{mplblanchedalmond}{rgb}{1.0,0.921568627451,0.803921568627}
\definecolor{mplblue}{rgb}{0.0,0.0,1.0}
\definecolor{mplblueviolet}{rgb}{0.541176470588,0.16862745098,0.886274509804}
\definecolor{mplbrown}{rgb}{0.647058823529,0.164705882353,0.164705882353}
\definecolor{mplburlywood}{rgb}{0.870588235294,0.721568627451,0.529411764706}
\definecolor{mplcadetblue}{rgb}{0.372549019608,0.619607843137,0.627450980392}
\definecolor{mplchartreuse}{rgb}{0.498039215686,1.0,0.0}
\definecolor{mplchocolate}{rgb}{0.823529411765,0.411764705882,0.117647058824}
\definecolor{mplcoral}{rgb}{1.0,0.498039215686,0.313725490196}
\definecolor{mplcornflowerblue}{rgb}{0.392156862745,0.58431372549,0.929411764706}
\definecolor{mplcornsilk}{rgb}{1.0,0.972549019608,0.862745098039}
\definecolor{mplcrimson}{rgb}{0.862745098039,0.078431372549,0.235294117647}
\definecolor{mplcyan}{rgb}{0.0,1.0,1.0}
\definecolor{mpldarkblue}{rgb}{0.0,0.0,0.545098039216}
\definecolor{mpldarkcyan}{rgb}{0.0,0.545098039216,0.545098039216}
\definecolor{mpldarkgoldenrod}{rgb}{0.721568627451,0.525490196078,0.043137254902}
\definecolor{mpldarkgray}{rgb}{0.662745098039,0.662745098039,0.662745098039}
\definecolor{mpldarkgreen}{rgb}{0.0,0.392156862745,0.0}
\definecolor{mpldarkgrey}{rgb}{0.662745098039,0.662745098039,0.662745098039}
\definecolor{mpldarkkhaki}{rgb}{0.741176470588,0.717647058824,0.419607843137}
\definecolor{mpldarkmagenta}{rgb}{0.545098039216,0.0,0.545098039216}
\definecolor{mpldarkolivegreen}{rgb}{0.333333333333,0.419607843137,0.18431372549}
\definecolor{mpldarkorange}{rgb}{1.0,0.549019607843,0.0}
\definecolor{mpldarkorchid}{rgb}{0.6,0.196078431373,0.8}
\definecolor{mpldarkred}{rgb}{0.545098039216,0.0,0.0}
\definecolor{mpldarksalmon}{rgb}{0.913725490196,0.588235294118,0.478431372549}
\definecolor{mpldarkseagreen}{rgb}{0.560784313725,0.737254901961,0.560784313725}
\definecolor{mpldarkslateblue}{rgb}{0.282352941176,0.239215686275,0.545098039216}
\definecolor{mpldarkslategray}{rgb}{0.18431372549,0.309803921569,0.309803921569}
\definecolor{mpldarkslategrey}{rgb}{0.18431372549,0.309803921569,0.309803921569}
\definecolor{mpldarkturquoise}{rgb}{0.0,0.807843137255,0.819607843137}
\definecolor{mpldarkviolet}{rgb}{0.580392156863,0.0,0.827450980392}
\definecolor{mpldeeppink}{rgb}{1.0,0.078431372549,0.576470588235}
\definecolor{mpldeepskyblue}{rgb}{0.0,0.749019607843,1.0}
\definecolor{mpldimgray}{rgb}{0.411764705882,0.411764705882,0.411764705882}
\definecolor{mpldimgrey}{rgb}{0.411764705882,0.411764705882,0.411764705882}
\definecolor{mpldodgerblue}{rgb}{0.117647058824,0.564705882353,1.0}
\definecolor{mplfirebrick}{rgb}{0.698039215686,0.133333333333,0.133333333333}
\definecolor{mplfloralwhite}{rgb}{1.0,0.980392156863,0.941176470588}
\definecolor{mplforestgreen}{rgb}{0.133333333333,0.545098039216,0.133333333333}
\definecolor{mplfuchsia}{rgb}{1.0,0.0,1.0}
\definecolor{mplgainsboro}{rgb}{0.862745098039,0.862745098039,0.862745098039}
\definecolor{mplghostwhite}{rgb}{0.972549019608,0.972549019608,1.0}
\definecolor{mplgold}{rgb}{1.0,0.843137254902,0.0}
\definecolor{mplgoldenrod}{rgb}{0.854901960784,0.647058823529,0.125490196078}
\definecolor{mplgray}{rgb}{0.501960784314,0.501960784314,0.501960784314}
\definecolor{mplgreen}{rgb}{0.0,0.501960784314,0.0}
\definecolor{mplgreenyellow}{rgb}{0.678431372549,1.0,0.18431372549}
\definecolor{mplgrey}{rgb}{0.501960784314,0.501960784314,0.501960784314}
\definecolor{mplhoneydew}{rgb}{0.941176470588,1.0,0.941176470588}
\definecolor{mplhotpink}{rgb}{1.0,0.411764705882,0.705882352941}
\definecolor{mplindianred}{rgb}{0.803921568627,0.360784313725,0.360784313725}
\definecolor{mplindigo}{rgb}{0.294117647059,0.0,0.509803921569}
\definecolor{mplivory}{rgb}{1.0,1.0,0.941176470588}
\definecolor{mplkhaki}{rgb}{0.941176470588,0.901960784314,0.549019607843}
\definecolor{mpllavender}{rgb}{0.901960784314,0.901960784314,0.980392156863}
\definecolor{mpllavenderblush}{rgb}{1.0,0.941176470588,0.960784313725}
\definecolor{mpllawngreen}{rgb}{0.486274509804,0.988235294118,0.0}
\definecolor{mpllemonchiffon}{rgb}{1.0,0.980392156863,0.803921568627}
\definecolor{mpllightblue}{rgb}{0.678431372549,0.847058823529,0.901960784314}
\definecolor{mpllightcoral}{rgb}{0.941176470588,0.501960784314,0.501960784314}
\definecolor{mpllightcyan}{rgb}{0.878431372549,1.0,1.0}
\definecolor{mpllightgoldenrodyellow}{rgb}{0.980392156863,0.980392156863,0.823529411765}
\definecolor{mpllightgreen}{rgb}{0.564705882353,0.933333333333,0.564705882353}
\definecolor{mpllightgrey}{rgb}{0.827450980392,0.827450980392,0.827450980392}
\definecolor{mpllightpink}{rgb}{1.0,0.713725490196,0.756862745098}
\definecolor{mpllightsalmon}{rgb}{1.0,0.627450980392,0.478431372549}
\definecolor{mpllightseagreen}{rgb}{0.125490196078,0.698039215686,0.666666666667}
\definecolor{mpllightskyblue}{rgb}{0.529411764706,0.807843137255,0.980392156863}
\definecolor{mpllightslategray}{rgb}{0.466666666667,0.533333333333,0.6}
\definecolor{mpllightslategrey}{rgb}{0.466666666667,0.533333333333,0.6}
\definecolor{mpllightsteelblue}{rgb}{0.690196078431,0.76862745098,0.870588235294}
\definecolor{mpllightyellow}{rgb}{1.0,1.0,0.878431372549}
\definecolor{mpllime}{rgb}{0.0,1.0,0.0}
\definecolor{mpllimegreen}{rgb}{0.196078431373,0.803921568627,0.196078431373}
\definecolor{mpllinen}{rgb}{0.980392156863,0.941176470588,0.901960784314}
\definecolor{mplmagenta}{rgb}{1.0,0.0,1.0}
\definecolor{mplmaroon}{rgb}{0.501960784314,0.0,0.0}
\definecolor{mplmediumaquamarine}{rgb}{0.4,0.803921568627,0.666666666667}
\definecolor{mplmediumblue}{rgb}{0.0,0.0,0.803921568627}
\definecolor{mplmediumorchid}{rgb}{0.729411764706,0.333333333333,0.827450980392}
\definecolor{mplmediumpurple}{rgb}{0.576470588235,0.439215686275,0.858823529412}
\definecolor{mplmediumseagreen}{rgb}{0.235294117647,0.701960784314,0.443137254902}
\definecolor{mplmediumslateblue}{rgb}{0.482352941176,0.407843137255,0.933333333333}
\definecolor{mplmediumspringgreen}{rgb}{0.0,0.980392156863,0.603921568627}
\definecolor{mplmediumturquoise}{rgb}{0.282352941176,0.819607843137,0.8}
\definecolor{mplmediumvioletred}{rgb}{0.780392156863,0.0823529411765,0.521568627451}
\definecolor{mplmidnightblue}{rgb}{0.0980392156863,0.0980392156863,0.439215686275}
\definecolor{mplmintcream}{rgb}{0.960784313725,1.0,0.980392156863}
\definecolor{mplmistyrose}{rgb}{1.0,0.894117647059,0.882352941176}
\definecolor{mplmoccasin}{rgb}{1.0,0.894117647059,0.709803921569}
\definecolor{mplnavajowhite}{rgb}{1.0,0.870588235294,0.678431372549}
\definecolor{mplnavy}{rgb}{0.0,0.0,0.501960784314}
\definecolor{mploldlace}{rgb}{0.992156862745,0.960784313725,0.901960784314}
\definecolor{mplolive}{rgb}{0.501960784314,0.501960784314,0.0}
\definecolor{mplolivedrab}{rgb}{0.419607843137,0.556862745098,0.137254901961}
\definecolor{mplorange}{rgb}{1.0,0.647058823529,0.0}
\definecolor{mplorangered}{rgb}{1.0,0.270588235294,0.0}
\definecolor{mplorchid}{rgb}{0.854901960784,0.439215686275,0.839215686275}
\definecolor{mplpalegoldenrod}{rgb}{0.933333333333,0.909803921569,0.666666666667}
\definecolor{mplpalegreen}{rgb}{0.596078431373,0.98431372549,0.596078431373}
\definecolor{mplpalevioletred}{rgb}{0.686274509804,0.933333333333,0.933333333333}
\definecolor{mplpapayawhip}{rgb}{1.0,0.937254901961,0.835294117647}
\definecolor{mplpeachpuff}{rgb}{1.0,0.854901960784,0.725490196078}
\definecolor{mplperu}{rgb}{0.803921568627,0.521568627451,0.247058823529}
\definecolor{mplpink}{rgb}{1.0,0.752941176471,0.796078431373}
\definecolor{mplplum}{rgb}{0.866666666667,0.627450980392,0.866666666667}
\definecolor{mplpowderblue}{rgb}{0.690196078431,0.878431372549,0.901960784314}
\definecolor{mplpurple}{rgb}{0.501960784314,0.0,0.501960784314}
\definecolor{mplred}{rgb}{1.0,0.0,0.0}
\definecolor{mplrosybrown}{rgb}{0.737254901961,0.560784313725,0.560784313725}
\definecolor{mplroyalblue}{rgb}{0.254901960784,0.411764705882,0.882352941176}
\definecolor{mplsaddlebrown}{rgb}{0.545098039216,0.270588235294,0.0745098039216}
\definecolor{mplsalmon}{rgb}{0.980392156863,0.501960784314,0.447058823529}
\definecolor{mplsandybrown}{rgb}{0.980392156863,0.643137254902,0.376470588235}
\definecolor{mplseagreen}{rgb}{0.180392156863,0.545098039216,0.341176470588}
\definecolor{mplseashell}{rgb}{1.0,0.960784313725,0.933333333333}
\definecolor{mplsienna}{rgb}{0.627450980392,0.321568627451,0.176470588235}
\definecolor{mplsilver}{rgb}{0.752941176471,0.752941176471,0.752941176471}
\definecolor{mplskyblue}{rgb}{0.529411764706,0.807843137255,0.921568627451}
\definecolor{mplslateblue}{rgb}{0.41568627451,0.352941176471,0.803921568627}
\definecolor{mplslategray}{rgb}{0.439215686275,0.501960784314,0.564705882353}
\definecolor{mplslategrey}{rgb}{0.439215686275,0.501960784314,0.564705882353}
\definecolor{mplsnow}{rgb}{1.0,0.980392156863,0.980392156863}
\definecolor{mplspringgreen}{rgb}{0.0,1.0,0.498039215686}
\definecolor{mplsteelblue}{rgb}{0.274509803922,0.509803921569,0.705882352941}
\definecolor{mpltab:blue}{rgb}{0.1216,0.4667,0.7059}
\definecolor{mpltan}{rgb}{0.823529411765,0.705882352941,0.549019607843}
\definecolor{mplteal}{rgb}{0.0,0.501960784314,0.501960784314}
\definecolor{mplthistle}{rgb}{0.847058823529,0.749019607843,0.847058823529}
\definecolor{mpltomato}{rgb}{1.0,0.388235294118,0.278431372549}
\definecolor{mplturquoise}{rgb}{0.250980392157,0.878431372549,0.81568627451}
\definecolor{mplviolet}{rgb}{0.933333333333,0.509803921569,0.933333333333}
\definecolor{mplwheat}{rgb}{0.960784313725,0.870588235294,0.701960784314}
\definecolor{mplwhite}{rgb}{1.0,1.0,1.0}
\definecolor{mplwhitesmoke}{rgb}{0.960784313725,0.960784313725,0.960784313725}
\definecolor{mplyellow}{rgb}{1.0,1.0,0.0}
\definecolor{mplyellowgreen}{rgb}{0.603921568627,0.803921568627,0.196078431373}

\usepackage{color,soul}
\usepackage{scalerel}
\usepackage{dashrule}
\usepackage{totcount}
\newtotcounter{citenum}
\def\oldcite{}
\let\oldcite=\bibcite
\def\bibcite{\stepcounter{citenum}\oldcite}
\usepackage{mathtools}
\usepackage[thinc]{esdiff}

\received{March 10, 2025}
\revised{May 6, 2024}
\accepted{May 19, 2025}
\published{\textit{The Astrophysical Journal Letters}}

\shorttitle{Where Have All the Little Red Dots Gone?}
\shortauthors{Khan et al.}

\begin{document}

\title{Where Have All the Little Red Dots Gone?\\
Supermassive Black Hole Binary Dynamics and its Impact on Galaxy Properties}

\author[0000-0002-5707-4268,gname=Fazeel,sname=Khan]{Fazeel Mahmood Khan}
\affiliation{New York University Abu Dhabi, P.O. Box 129188, Abu Dhabi, United Arab Emirates}
\affiliation{Center for Astrophysics and Space Science (CASS), New York University Abu Dhabi, PO Box 129188, Abu Dhabi, UAE}
\email[show]{\href{mailto:fmk5060@nyu.edu}{fmk5060@nyu.edu}}

\author[0000-0002-4306-5950,gname=Benjamin,sname=Davis]{Benjamin L.\ Davis}
\affiliation{New York University Abu Dhabi, P.O. Box 129188, Abu Dhabi, United Arab Emirates}
\affiliation{Center for Astrophysics and Space Science (CASS), New York University Abu Dhabi, PO Box 129188, Abu Dhabi, UAE}
\email{ben.davis@nyu.edu}

\author[0000-0002-8171-6507,gname=Andrea,sname=Macci\`{o}]{Andrea Valerio Macci\`{o}}
\affiliation{New York University Abu Dhabi, P.O. Box 129188, Abu Dhabi, United Arab Emirates}
\affiliation{Center for Astrophysics and Space Science (CASS), New York University Abu Dhabi, PO Box 129188, Abu Dhabi, UAE}
\affiliation{Max-Planck-Institut f\"{u}r Astronomie, K\"{o}nigstuhl 17, 69117 Heidelberg, Germany}
\email{maccio@nyu.edu}

\author[0000-0003-2227-1322,gname=Kelly,sname=Holley-Bockelmann]{Kelly Holley-Bockelmann}
\affiliation{Department of Physics and Astronomy, Vanderbilt University, Nashville, TN 37240, USA}
\affiliation{Department of Physics, Fisk University, Nashville, TN 37208, USA}
\email{k.holley@vanderbilt.edu}

\begin{abstract}

Recent \textit{James Webb Space Telescope} observations have revealed a peculiar class of galaxies at redshifts $z \gtrsim 6$, characterized by extremely high central stellar densities and overmassive central supermassive black holes (SMBHs), ``little red dots'' (LRDs).
A critical question remains: If LRDs were common at high redshifts, how would they evolve into local elliptical galaxies with significantly lower central densities?
To address this, we performed direct $N$-body simulations of LRD mergers, focusing on the coevolution of host galaxies and central SMBHs.
We track the complete evolution of SMBH binaries into the three-body hardening and gravitational-wave (GW) emission phase.
Our results demonstrate that during galaxy mergers, the central SMBHs can eject a substantial amount of mass from the galactic core via the three-body slingshot effect, leading to a decrease in central stellar surface density by an order of magnitude.
Additionally, GW recoil can further contribute in making the galaxy centers less dense and more in alignment with low-redshift quiescent galaxies. 
This transformation occurs on a relatively short timescale of a few $\sim$100\,Myr, implying that LRDs can evolve into lower-redshift elliptical galaxies by $z<4$.
The timescales for our SMBH mergers vary between 100\,Myr and 800\,Myr, depending on the initial orbital parameters of the merging galaxies and the mass ratio of the SMBHs.
Our findings provide a plausible mechanism for the transformation of LRDs into elliptical galaxies while highlighting the efficiency of SMBH mergers in such high-density environments, which plays a crucial role in SMBH growth.

\end{abstract}

\keywords{
\href{http://astrothesaurus.org/uat/159}{Black hole physics (159)};
\href{http://astrothesaurus.org/uat/573}{Galaxies (573)};
\href{http://astrothesaurus.org/uat/591}{Galaxy dynamics (591)};
\href{http://astrothesaurus.org/uat/594}{Galaxy evolution (594)};
\href{http://astrothesaurus.org/uat/602}{Galaxy kinematics (602)};
\href{http://astrothesaurus.org/uat/609}{Galaxy nuclei (609)};
\href{http://astrothesaurus.org/uat/612}{Galaxy physics (612)};
\href{http://astrothesaurus.org/uat/615}{Galaxy properties (615)};
\href{http://astrothesaurus.org/uat/678}{Gravitational waves (678)};
\href{http://astrothesaurus.org/uat/1083}{$N$-body simulations (1083)};
\href{http://astrothesaurus.org/uat/1663}{Supermassive black holes (1663)}
}

\section{Introduction}\label{sec:intro}

Supermassive black holes (SMBHs) are key to many physical processes happening in galaxy centers.
Their presence in galactic nuclei is related with many correlations such as the black hole mass scaling relations, $M_\bullet$--$\sigma$ \citep{Ferrarese:2000,Gebhardt:2000,Davis:2017,Sahu:2019b} and $M_\bullet$--$M_\mathrm{bulge}$ \citep{Magorrian:1998,Marconi:2003,Haring:2004,Davis:2019,Sahu:2019,Graham:2023}, between the central velocity dispersion and bulge (spheroid) mass, respectively.
Recent \textit{James Webb Space Telescope} (\textit{JWST}) observations at high redshift reveal that SMBHs may contain a large fraction of galaxy stellar mass, roughly 10\% or greater \citep{Goulding:2023,Kocevski:2023,Bogdan:2024,Marshall:2024,Natarajan:2024}.
This is about 2--3 orders of magnitude higher than what local scaling relations predict \citep{Davis:2018,Sahu:2019}.\footnote{
However, \citet{Li:2025} demonstrate that the alleged overmassive black holes found by the \textit{JWST} are not inconsistent with local scaling relations after accounting for the coupled effects of measurement uncertainties and selection biases.
Still, \citet{Sun:2025} find a significant increase over the local scaling relations in very low-mass galaxies ($<$$10^{10}\,\mathrm{M}_\sun$) for $z\geq4$.
}
SMBH hosts observed at high redshift ($z \gtrsim 4$) by the \textit{JWST} are often red in color and compact in size, leading to their sobriquet ``little red dots'' \citep[LRDs;][]{Matthee:2024,Greene:2024,Williams:2024,Ananna:2024,Yue:2024,Kokubo:2024,Li:2024,Volonteri:2024,Akins:2025,Furtak:2025}.

The defining characteristics of LRDs, their compactness and their overmassive black holes, are in strong tension with our understanding of the coevolution of SMBHs and galaxies across cosmic time.
\citet{Lin:2024} suggest that low-redshift ($z<0.4$) ``green pea'' galaxies hosting broad-line active galactic nuclei (AGNs) with overmassive black holes are local analogs of LRDs.
\citet{Bellovary:2025} hypothesizes that ``little red dots are tidal disruption events in runaway-collapsing clusters.''
\citet{Inayoshi:2025} presumes that the characteristic features of LRDs naturally fade with subsequent AGN episodes.

\added{
The LRDs with their high stellar masses ($10^{10}$--$10^{11}\,\mathrm{M}_\odot$) observed at redshifts $z \sim 6$--8 are expected to continue growing over cosmic time and are likely progenitors of massive elliptical galaxies observed at $z = 0$.
Notably, the  number density of massive ellipticals ($\sim$$10^{-3}$--$10^{-5}\,\mathrm{Mpc}^{-3}$) at $z = 0$  \citep{Ilbert2013,vanderwel2014,Bernardi2017} fares well within the observed range of the comoving number density of LRDs  ($\sim$$10^{-4}\,\mathrm{Mpc}^{-3}$) at $z>4$ prior to a steep decline \citep{Weibel:2024,Pablo:2024,Ma:2025}, supporting this evolutionary sequence.

LRDs also share obvious similarities with ``red nugget'' galaxies, i.e., compact, massive ($10^{10}$--$10^{11}\,\mathrm{M}_\sun$), quiescent galaxies observed at redshifts $z\sim1$--2 \citep{Damjanov:2009}.
It is conceivable that a subset of LRDs---particularly those with AGN-driven quenching---could evolve into red nuggets.
\citet{Rantala2024} suggested that red nuggets later on can transform into high-mass ellipticals through a series of galaxy and SMBH mergers. 
Likewise, ancient LRDs could possibly be masquerading as the extant bulges/spheroids of larger galaxies at low redshifts.
Specifically, \citet{Hon:2022} describe a ``disk cloaking'' scenario whereby red nuggets (at $1\lesssim z\lesssim2$) develop a rotational disk via mergers and accretion, thus cloaking red nuggets as the classical bulges of disk galaxies in the local Universe.
}

If LRDs are extremely compact galaxies as the observations suggest, a key puzzle remains: How do they transform into local elliptical galaxies, especially since their dense cores need to be erased by some mechanism efficiently so that their densities drop considerably by the time they reach $z=4$?
There are two mechanisms that could potentially lead to an expansion of the central stellar density:
\begin{enumerate}
    \item Non-adiabatic collisionless expansion of the central stellar distribution in response to gas ejection by AGNs feedback \citep[e.g.,][]{Teyssier2011,Arora2024}.
    \item Core scouring by the SMBH binary in the three-body scattering phase following the merger of two LRDs, each with their own overmassive SMBH \citep[e.g.,][]{graham+04,mer06}.
\end{enumerate}
If LRDs are not dominated by AGNs, as most studies suggest \citep{Ananna:2024,Yue:2024,Kokubo:2024}, then mechanism~\#1 would not be very efficient.
On the other hand, overmassive SMBHs suspected to be present in the center of these galaxies can drive significant mass out of the centers when they inspiral \citep{khan+12a,Nasim2021}, reducing the central stellar densities via mechanism~\#2, and bringing them closer to the observed values in low-$z$ ellipticals.
However, it remains to be seen how much time a binary spends in a hard binary phase, where it loses its energy via three-body scattering with stars on centrophilic orbits \citep{merritt2004,Baile15}.
In a previous study, \citep{Khan+15} examined binary dynamics involving overmassive black holes accounting for 20--60\% of their host galaxy's mass and demonstrated that these overmassive black holes merge exceptionally quickly.
This rapid merger process nearly bypasses the three-body scattering phase, resulting in minimal disruption to the central stellar density of the merger remnant.

In this study, we simulate mergers of two LRDs, systematically varying parameters such as their masses, the masses of their SMBHs, and the mass ratios of both the merging galaxies and their SMBHs.
Our investigation aims to address two key questions: 
\begin{itemize}
    \item How much mass is carved out by the massive binary, and what do the product galaxies look like in terms of stellar density after accommodating an SMBH merger? 
    \item Following the merger of two LRDs, how quickly do SMBHs evolve in the merger remnant, and do they achieve a coalescence followed by gravitational-wave (GW) emission?
\end{itemize}
We begin by detailing the numerical simulation setup (\S\ref{sec:setup}), then we provide the results of our SMBH binary dynamics (\S\ref{sec:dynamics}) and the properties of the merging galaxy systems (\S\ref{sec:properties}), we engage in a discussion to interpret and summarize our results (\S\ref{sec:discussion}), and finally we leave some parting conclusions (\S\ref{sec:conclusions}).

\section{Numerical Setup}\label{sec:setup}

We build galaxy models with parameters motivated by LRDs.
In particular, we used two models with structural parameters from \citet{Baggen:2024} with RUBIES IDs: 49140 and 55604, measured with spectroscopic redshifts of 6.68 and 6.98, respectively.
The authors present three estimates for each galaxy's stellar mass depending on the contribution from AGNs.
The minimal (or no contribution) model has the maximum stellar mass, moderate contribution results in an intermediate mass, and maximum contribution of AGNs cause the stellar mass to drop to the minimum value.

\citet{Baggen:2024} fit all light profiles with \citet{Sersic:1963} profiles that have indices of $n=1.5$.
These galaxies are very compact systems with effective radii on the order of 100\,pc with large uncertainties.\footnote{
These extremely small sizes for $z=7$ galaxies are considerably smaller than simulations predicted prior to the launch of \text{JWST} \citep{Marshall:2020}.
}
We adopted the moderate mass model with a conservative effective radius value of 100\,pc for our reference models.
The SMBH masses are adopted from \citet{Wang:2024}, for the intermediate galaxy mass scenario.
For comparison, the stellar masses of the LRDs analyzed by recent works \citep{Labbe:2025,Kocevski:2024,Xiao:2025} are consistent with our models.
Furthermore, there is emerging evidence of precursors to LRD mergers; \citet{Tanaka:2024} and \citet{Rosa2025} discovered dual LRD candidates.

\begin{figure}
\centerline{
  \resizebox{\hsize}{!}{\includegraphics{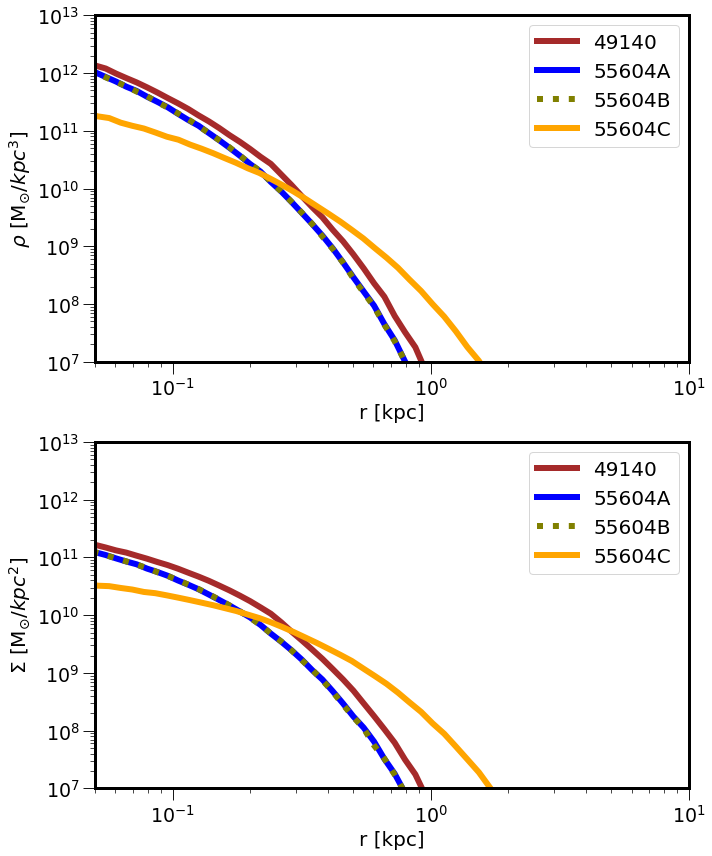}}
  }
\caption{
Initial volume density ($\rho$) profiles of our galaxy models (top panel) and surface stellar density ($\Sigma_\star$) profiles of the same (bottom panel).
} \label{fig:densini}
\end{figure}

Table~\ref{tab:simparam} shows the key parameters for our initial galaxy models derived from the properties of RUBIES.
Figure~\ref{fig:densini} shows the initial volume (top panel) and surface density profiles (bottom panel) of our galaxy models. 
Models 49140M and 55604M are not plotted because they have exactly the same profiles as galaxies 49140 and 55604A, respectively.
The only difference lies in the mass of their central SMBHs, which are 10\% of their host galaxy masses.
The model with a larger effective radius has both a lower volume and a lower surface density.

\begin{deluxetable}{lrrcrc}
\tablecolumns{6}
\tablecaption{Initial Galaxy Parameters}\label{tab:simparam}
\tablehead{
Model & $M_\bullet$ & $M_\star$ & $R_\mathrm{e}$ & $N_{\rm part}$ & $b/a,c/a$ \\
 & ($10^7$\,M$_{\odot}$) & ($10^9$\,M$_{\odot}$) & (pc) & ($10^5$) & \\
\colhead{(1)} & \colhead{(2)} & \colhead{(3)} & \colhead{(4)} & \colhead{(5)} & \colhead{(6)}
}
\startdata
49140 & 50 & 10.0 & 123 & 10 & $0.9,0.60$ \\
49140M & 100 & 10.0 & 123 & 10 & $0.9,0.60$ \\
55604A & 30 & 6.2 & 100 & 6 & $0.9,0.75$ \\
55604B & 3 & 6.2 & 100 & 6 & $0.9,0.75$ \\
55604C & 30 & 6.2 & 300 & 6 & $0.9,0.75$ \\
55604M & 60 & 6.2 & 100 & 6 & $0.9,0.75$ \\
\enddata
\tablecomments{
Column~(1): Galaxy model name.
Column~(2): SMBH mass (in M$_{\odot}$).
Column~(3): Stellar mass (in M$_{\odot}$).
Column~(4): Effective radius (in pc).
Column~(5): Number of particles.
Column~(6): Axes ratios.
}
\end{deluxetable}

Table~\ref{tab:mergerparam} presents the parameters and progenitors of our galaxy merger runs.
We explore the effect of different impact parameters of the orbits, and mass ratios of merging galaxies and SMBHs in our runs. 
We placed the two galaxies on eccentric orbits at an initial apocenter separation of 1\,kpc, which is roughly 10 times the effective radius of the progenitor galaxies.
The pericenter approach is set to be 100\,pc for all runs except for run M2, where two centers are about 500\,pc away during their first closest approach.

\begin{deluxetable}{lcccccccc}
\tablecolumns{8}
\tablefontsize{\scriptsize}
\tablecaption{Galaxy Merger Parameters}\label{tab:mergerparam}
\tablehead{
Run & Gal1 & Gal2 & $b$ & $q_\mathrm{gal}$ & $q_\bullet$ & $s$ & $T_\mathrm{merge}$ \\
 & & & (pc) & & & $\left ( \frac{\mathrm{pc}^{-1}}{\mathrm{Myr}}\right )$ & (Myr) \\
\colhead{(1)} & \colhead{(2)} & \colhead{(3)} & \colhead{(4)} & \colhead{(5)} & \colhead{(6)} & \colhead{(7)} & \colhead{(8)}
}
\startdata
M1 & 49140 & 55604A & 200 & $0.6$ & $0.6$ & 0.061 & 393 \\
M2 & 49140 & 55604A & 500 & $0.6$ & $0.6$ & 0.021 & 794 \\
M3 & 49140 & 55604B & 200 & $0.6$ & $0.06$ & 0.152 & 125 \\
M4 & 49140 & 55604C & 200 & $0.6$ & $0.6$ & 0.035 & 428 \\
M5 & 49140M & 55604M & 200 & $0.6$ & $0.6$ & 0.019 & 532 \\
\enddata
\tablecomments{
Column~(1): Galaxy merger run.
Column~(2): Primary galaxy.
Column~(3): Secondary galaxy.
Column~(4): Impact parameter (in pc).
Column~(5): Secondary-to-primary galaxy mass ratio.
Column~(6): Secondary-to-primary SMBH mass ratio.
Column~(7): The hardening rate (in pc$^{-1}$\,Myr$^{-1}$).
Column~(8): Merger time for SMBHs (in Myr).
}
\end{deluxetable}

We use direct $N$-body code $\phi$-GPU \citep{Berczik:2011,Berczik:2013} to run our galaxy merger simulations.
For pairwise forces, we applied 0.1\,pc softening between stellar particle interactions and 0.001\,pc for star--SMBH interactions, while SMBH--SMBH interactions are calculated with zero softening.
We incorporate post-Newtonian (PN) terms up to 2.5PN order in the equations of motion for the SMBH binary to accommodate relativistic effects.
In particular, energy loss by gravitational waves is taken care of by 2.5PN terms \citep{blanchet2006}.

\section{SMBH Binary Dynamics}\label{sec:dynamics}

\begin{figure*}
\centerline{
  \resizebox{\hsize}{!}{\includegraphics{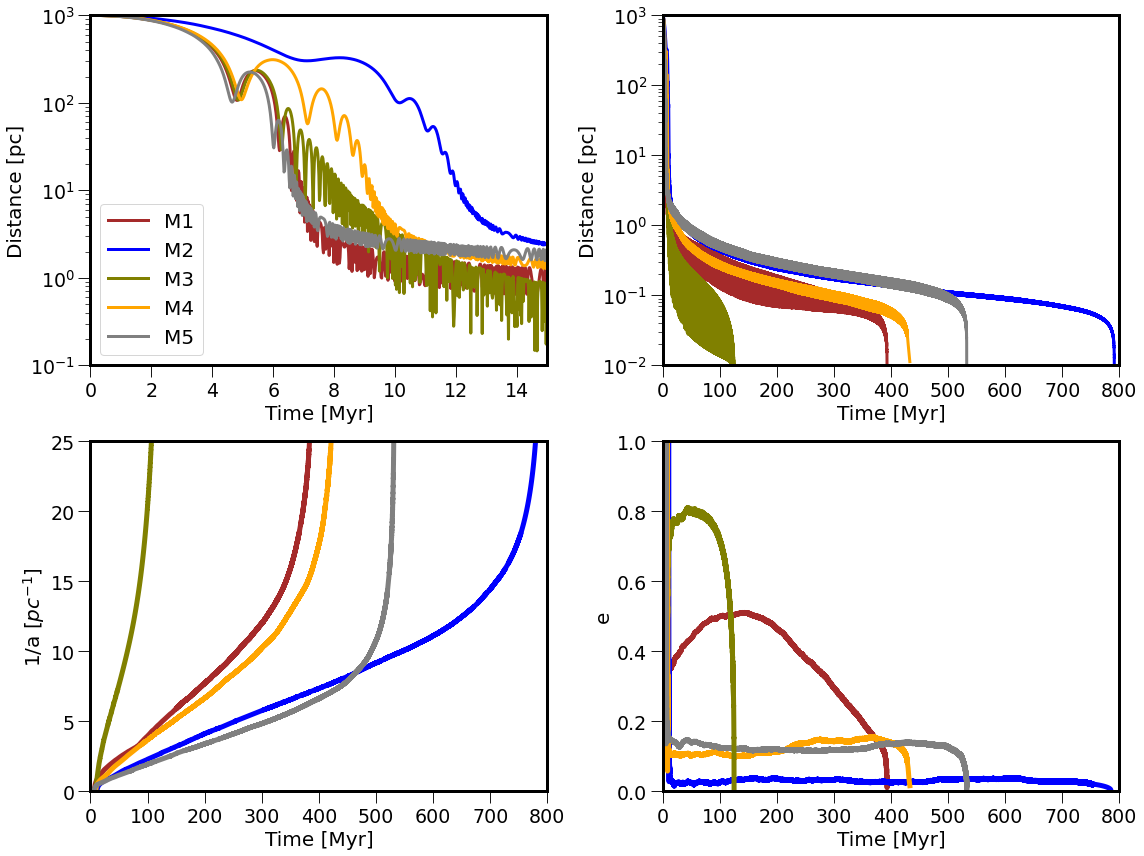}}
  }
\caption{
SMBH binary parameters for our galaxy merger study with five different simulation runs.
Top left panel: separation of the pair of SMBHs as the galaxies merge during the initial phase of their runs.
Top right panel: the complete evolution of the separation between SMBHs as they pair and merge, driven by dynamical friction, $3-$body scattering and gravitational-wave emission.
Bottom left panel: the inverse semimajor axis of each SMBH binary system as a function of time. 
Bottom right panel: the eccentricity evolution of the SMBH binaries.
} \label{fig:SMBHparams}
\end{figure*}

We show a detailed analysis of all our runs in the four-panel plot presented in Figure~\ref{fig:SMBHparams}.

\subsection{SMBH Separation During Galaxy Mergers}

During the initial phase of galaxy mergers, the SMBH separations follow the relative distance between the centers of the two galaxies (Figure~\ref{fig:SMBHparams}, top left panel).
We have chosen two sets of orbits, four of our runs (namely, M1, M3, M4, and M5) follow an eccentric orbit where the initial pericenter of the merging galaxies is close to 100\,pc, which is similar to the effective radii of the merging galaxies.
We notice that after the first pericenter passage, the orbits shrink quickly as the galaxies lose energy due to dynamical friction and violent relaxation, and the galaxies merge subsequently after a few pericenter passages.

For M4, where the secondary galaxy had an effective radius that was three times larger, we witness a relatively prolonged galaxy merger time.
For run M2, we increased the impact parameter such that the two merging galaxies have a first pericenter separation of 500\,pc, which is well outside the effective radii of the merging galaxies, which slows down the energy loss, and hence prolongs the galaxy merger time.
Overall, the galaxy merger times range from 7--12\,Myr.

\subsection{SMBH Separation Until Their Coalescence}

The complete evolution of the separation between the SMBHs during the galaxy mergers and until the eventual coalescence of the SMBHs in our merger runs is presented in the top right panel of Figure~\ref{fig:SMBHparams}.
In this extension to the range presented in the top left panel, the simulation runs begin to distinguish themselves through stellar and post-Newtonian hardening. 

\subsection{Inverse Semimajor Axis Evolution}

The bottom left panel of Figure~\ref{fig:SMBHparams} traces the complete evolution of the inverse semimajor axis ($1/a$) of each SMBH binary system (a measure of the SMBH binary binding energy) once the two SMBHs are bound.
We notice different tracks for the $1/a$ evolution of the SMBH binaries in our runs.
We fit straight lines to $1/a$ in the linear regime to determine the hardening rate, $s=d(1/a)/dt$.
Table~\ref{fig:SMBHparams} shows the hardening rates for all our models.
Model~M3 has the highest hardening rate ($s = 0.152$\,pc$^{-1}$\,Myr$^{-1}$) because it has the least massive secondary and it is easier to extract the energy from less massive binaries.
The binary SMBH in the M3 run also has the highest eccentricity, which further helps the SMBHs to merge in a relatively short time span, $\sim$100\,Myr.

For the case of SMBHs with comparable masses, run M1 shows the highest hardening rate ($s = 0.0613$\,pc$^{-1}$\,Myr$^{-1}$), and also a relatively high eccentricity, resulting in a faster merger time.
For run M4, we noticed a lower hardening rate ($s=0.035$\,pc$^{-1}$\,Myr$^{-1}$), together with a lower value of eccentricity that results in a longer merger time for its SMBHs.
The binary SMBH in the M2 run has the lowest value of $s=0.021$\,pc$^{-1}$\,Myr$^{-1}$ for cases where the mass of the SMBH is 5\% of the total mass of the galaxy and is the run with the least eccentric orbit among the merging galaxy runs.
The lowest hardening rate combined with the lowest value of eccentricity means that the SMBH binary takes the longest to merge in the M2 merger.
The run M5 hosts the binary with the most massive SMBHs and the lowest value of $s=0.019$\,pc$^{-1}$\,Myr$^{-1}$. 

\subsection{Eccentricity Evolution}

The eccentricities of the SMBH binaries in our galaxy merger runs are presented in the bottom right panel of Figure~\ref{fig:SMBHparams}.
Runs M1, M3, M4, and M5 initially had more eccentric orbits, whereas run M2 had a less eccentric orbit for merging galaxies.
We notice that in run M1, the SMBH binary forms with a mild eccentricity ($e \sim 0.4$), which grows slightly and attains a value of $e \sim 0.55$ in the hard binary regime before gravitational waves take over in the final phase, resulting in the circularization of the orbit.
Run M2, which had a less eccentric orbit for merging galaxies, hosts an SMBH binary that is close to circular.

For run M3, where the secondary SMBH is 16 times less massive than the primary one, we observe a high eccentricity ($e \sim 0.8$) for the binary.
A high eccentricity value for the cases where the binary has a high mass ratio between SMBHs has been previously reported both in isolated galaxy models and in galaxy mergers \citep{Iwasawa+11,Khan+15}.
Finally, SMBH binaries form and retain a low eccentricity of $e \sim 0.1$ for runs M4 and M5.
This eccentricity behavior is consistent with our earlier findings, in which we witness low eccentricity binaries in mergers of dense systems \citep{khan+12a,khan18}.

\subsection{Gravitational-wave Recoil}

During the final phase of their mergers in a binary system, SMBHs emit gravitational waves asymmetrically, resulting in recoil of the coalesced SMBH.
The recoil velocities can reach up to 5,000\,km/s \citep{lusato25} for maximally spinning and mis-aligned SMBHs. 
Non-spinning SMBHs tend to recoil with a smaller velocity of $\sim$100\,km/s. 
It has been shown in earlier studies that SMBH recoils can contribute to core formation in low-$z$ ellipticals \citep{Khonji:2024,Rawlings:2025}.

Following the coalescence of SMBHs in our simulations, we place the SMBH with the combined mass of merging black holes in the binary at the binary center of mass and give it a recoil velocity in the $xy$-plane with a moderate value of 600\,km/s.
We evolve the system further in time with the recoiling SMBH until it settles back at the center assisted by dynamical friction. 
Here, this process takes about 10--30\,Myr. 

\section{Properties of the Merging Galaxy Systems}\label{sec:properties}

\begin{figure*}
\centerline{
  \resizebox{\hsize}{!}{\includegraphics{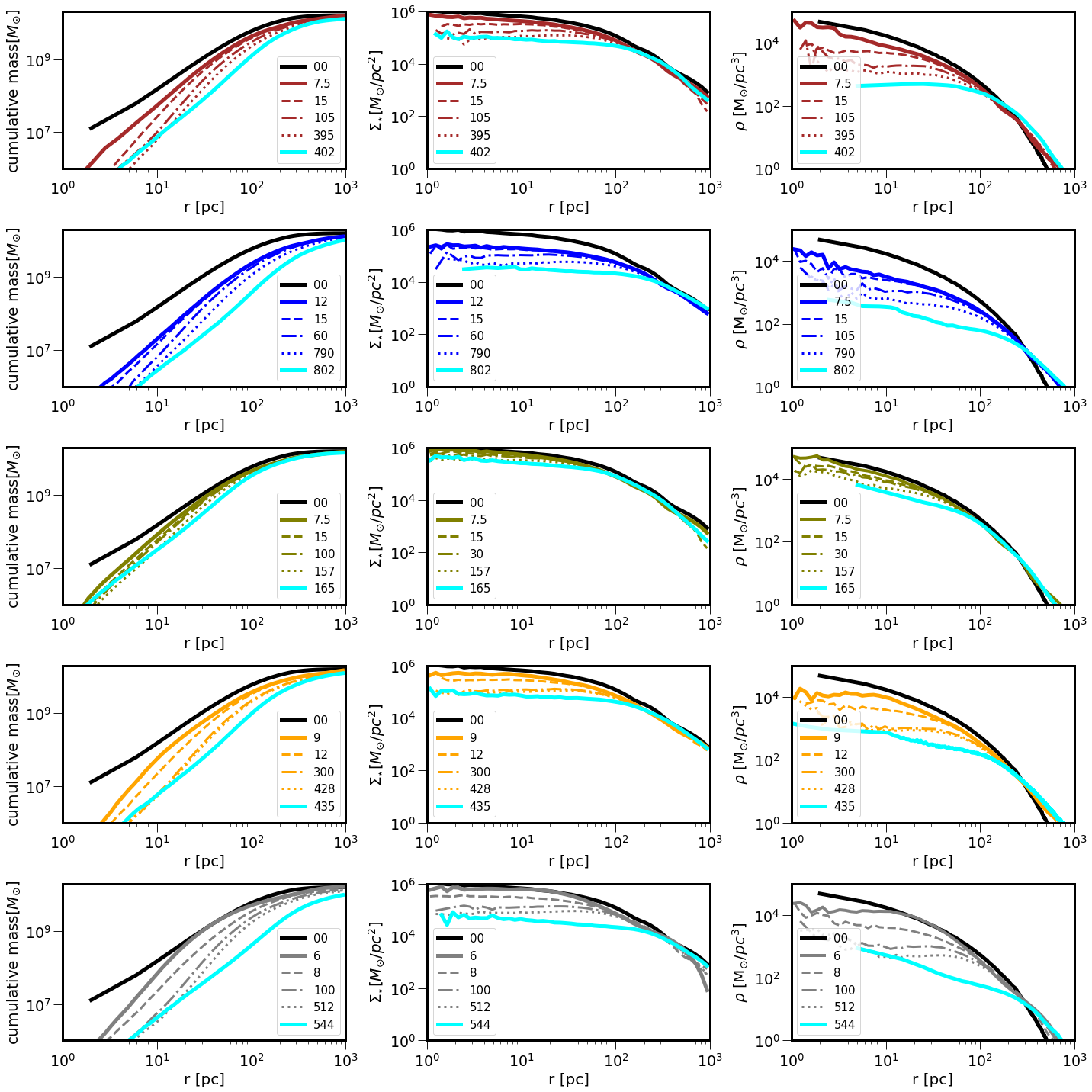}}
  }
\caption{
Evolution of galaxy profiles in merger products (from the top to bottom rows: M1, M2, M3, M4, and M5) across the times of binary formation, hard binary formation, and gravitational wave-dominated phases at the time of SMBH coalescence.
Profiles include cumulative mass (left column), surface stellar density ($\Sigma_\star$, middle column), and volume density ($\rho$, right column) at selected times (in Myr).
The start of each simulation is depicted as a black line (\textcolor{black}{\hdashrule[0.35ex]{8mm}{1pt}{}}) and the gravitational wave kick is always represented with a cyan line (\textcolor{cyan}{\hdashrule[0.35ex]{8mm}{1pt}{}}).
}\label{fig:mergerprofils}
\end{figure*}

In Figure~\ref{fig:mergerprofils}, we explore the evolution in the profiles of the newly formed galaxies following their mergers.
For reference, we also consider the $T=0$ (steady-state) profiles of the primary progenitor galaxy (black lines).
The left column presents the cumulative mass profile of the merged system approximately from the time of the SMBH binary formation to the final coalescence of the SMBHs.  
We notice a significant drop in mass in the central few hundred parsecs for all our merger models except for M3, where the secondary is the least massive and also merges in a relatively shorter time span of about 100\,Myr.

The middle and right columns display the projected ($\Sigma_\star$) and volume ($\rho$) densities, respectively.
We notice that $\Sigma_\star$ drops by almost an order of magnitude from $10^6\,\mathrm{M}_{\odot}$\,pc$^{-2}$ to $10^5 \,\mathrm{M}_{\odot}$\,pc$^{-2}$ in all cases, except when the secondary SMBH is 16 times less massive than the primary.
However, for this specific case, the evolution is minimal because of two primary reasons: (i) the smaller mass of the secondary makes it less efficient at core scouring, and (ii) the SMBH binary has a shorter coalescence time as a result of its higher eccentricity.
The volume density profiles follow a very similar trend with more prominent flattening toward the center when compared to that of the surface density.

\begin{figure}
\centerline{
  \resizebox{\hsize}{!}{\includegraphics{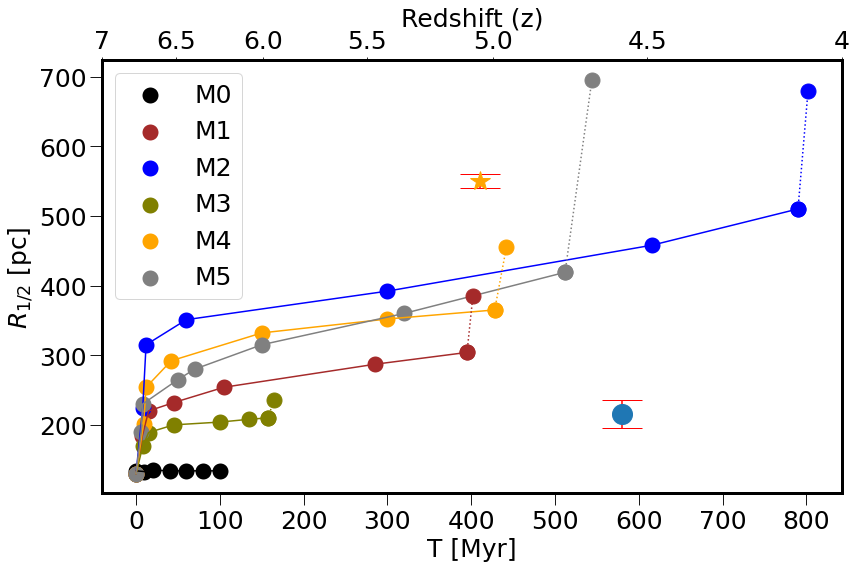}}
  }
\centerline{
  \resizebox{\hsize}{!}{\includegraphics{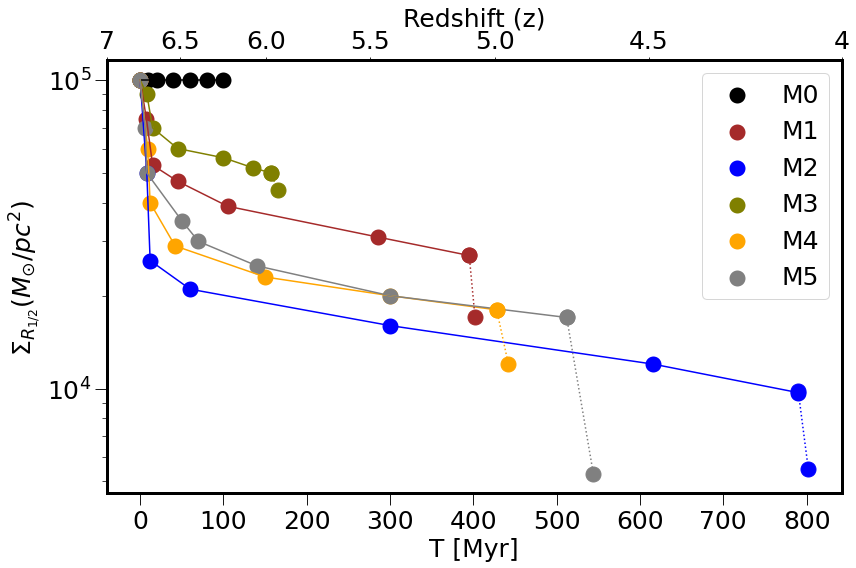}}
  }
\caption{
Time/redshift evolution of half-mass radii and surface densities within the half-mass radius for our various models.
The dotted lines always represent the evolution following the GW kick.
Top: time ($T$) evolution of the half-mass radii ($R_{1/2}$) of the newly formed galaxies resulting from mergers.
The half-mass radii increase as the SMBHs scour the central regions of their coalesced host galaxies and deposit mass into larger radii.
M0 is our stability run, and we witness that the half-mass radii at $T\sim100$\,Myr remains practically unchanged.
The \textcolor{mpltab:blue}{$\bullet$} \citep{Carnall23} and \textcolor{mplorange}{$\star$} \citep{deGraaf24} symbols depict the observed half-light radii for quiescent galaxies with masses similar to those used for our simulations.
Bottom: the surface density within the half-mass radius ($\Sigma_{R_{1/2}}$) plotted as a function of time.
}\label{fig:halfmass-radii}
\end{figure}

As a proxy of effective radius, we plotted how the (three-dimensional) half-mass radius ($R_{1/2}$) changes as SMBH binaries remove mass from the center of their host galaxy (see the top panel of Figure~\ref{fig:halfmass-radii}).
We notice an abrupt increase in $R_{1/2}$ around the binary formation time when the SMBH binary slingshots stellar particles, interacting with them on a dynamical timescale.
As the immediate region around the SMBH is largely scoured, we see a familiar mass exodus during the hard binary phase by centrophilic orbits \citep{Miralda-Escude:1989,khan+11,Gualandris+12,rantala+18}, resulting in a gradual increase of $R_{1/2}$.
The longer an SMBH binary spends in the 3-body scattering stage, the more damage it exerts.

As the SMBHs merge and the remnant SMBH recoils, we witness an additional increase in $R_{1/2}$ that ranges from 100--300\,pc, except for the M3 run, which has a minor merger of the SMBHs.
In our simulations, recoiling SMBHs drive the expansion of $R_{1/2}$ on relatively short timescales of a few tens of Myr as they settle back to the centers, facilitated by the high densities of their host galaxies.
This is in contrast to the Gyr recoil timescales observed in low-$z$ ellipticals.
We note that the total increase in $R_{1/2}$ caused by the three-body scattering and SMBH recoil is from a factor of two to a factor of seven, where the former case is for the least massive secondary (run M3).
The blue circle \citep{Carnall23} and the orange star \citep{deGraaf24} in Figure~\ref{fig:halfmass-radii} show the observed $R_{1/2}$ for quiescent galaxies at lower redshifts with masses similar to those used for LRDs in our studies.

We estimate the stellar surface density as $\Sigma_{R_{1/2}} = M_{\star}/\pi R_{1/2}^2 $, where $M_{\star}$ is the stellar mass enclosed within the half-mass radius, as shown in the bottom panel of Figure~\ref{fig:halfmass-radii}.
Note the considerable drop in surface density, by a factor of 3--10, for comparable mass ratio SMBH binaries.
These changes mirror the changes shown for the half-mass radii in the top panel.
When considering the additional effect of GW recoil, we see both runs M2 and M5 decrease their initial densities by more than an order of magnitude in much less than a Gyr of evolution.
If we compare these sizes and densities to the $M_\bullet$--$R_{1/2}$ \citep[][Fig.~12]{Sahu:2020} and $M_\bullet$--$\Sigma_{R_{1/2}}$ \citep[][Fig.~6]{Sahu:2022b} relations for local elliptical galaxies, we find that all our simulations start with very overmassive black holes; core scouring brings the galaxies closer to the local relation, but still remain overmassive.

\section{Discussion}\label{sec:discussion}

In this study, we performed a suite of galaxy merger simulations to investigate the coevolution of LRDs and their central SMBHs.
Following the galaxy merger, we evolved the SMBH binary evolution until coalescence by taking into account post-Newtonian gravitational terms up to the 2.5 order (the dominant gravitational wave term) for the binary equation of motion in the relativistic regime.
As the SMBHs binaries coalesced, we gave the remnant SMBH a moderate recoil velocity and evolved the system until it settled in the galactic center.
We analyzed the impact of SMBH binary evolution and recoil on the host galaxies by analyzing the merger remnant profiles.
We demonstrated that SMBH dynamics provide a robust mechanism for transforming high-redshift ($z\gtrsim6$) galaxies into local elliptical galaxies with lower central densities.
\added{Our findings are in line with the prevailing body of work that reflects redshift-dependent changes in stellar density and effective radius, consistent with inside-out growth \citep{Dokkum:2008,Bezanson:2009,Baggen:2023}.}
In the following, we discuss and summarize our key findings.

\subsection{The Impact of the SMBH Binary and Gravitational Wave Recoil on Host Galaxies}

The primary result of our study reveals that core scouring is a powerful mechanism to catalyze the evolution of LRDs into lower-redshift elliptical galaxies. 
Our simulations demonstrate that the three-body slingshot effect during the inspiral of a SMBH binary into the dense central environment of LRDs, coupled with SMBH recoil induced by gravitational wave emission, will naturally eject a considerable amount of mass from the galactic core and lead to a noticeable decrease in central stellar density by an order of magnitude.
Together, these processes increase the half-mass radius by a factor of 3--7 and reduce the central stellar surface density by an order of magnitude.
The timescale of this transformation is typically $\sim$400--800\,Myr, which allows LRDs to transform into low-$z$ elliptical galaxies by $z=4$--$5$ ($\sim$800\,Myr).

\citet{Rantala2024} performed numerical simulations to study the evolution of low-$z$ red nuggets and explored the effect of SMBH binary core scouring in a series of galaxy mergers.
They came up with a fitting relation which suggests that each SMBH merger can cause a 10\% increase in the host galaxy influence radius.
This discrepancy can be caused by two factors: 
\begin{enumerate}
    \item The galaxies in \citet{Rantala2024} have typical effective radii of 1\,kpc and SMBH binary slingshots would typically deposit the stars that interacted with the binary to within the same spatial extent. 
    Thus, only a few strong encounters in the hard binary phase would result in scattering at larger than 1\,kpc, resulting in a small increase in the effective or half-mass radii.
    Whereas, LRDs are very compact and hence three-body scattering events depositing the stars to a kpc scale would increase the initial half-mass radius of 100\,pc to a manifold of sizes.
    \item Additionally, our models have overmassive SMBHs, and binaries with larger masses have larger orbital velocities (for a given semimajor axis), resulting in stronger three-body slingshot kicks.
\end{enumerate}

The coalescence time of SMBH binaries in our merger simulations range from approximately 100\,Myr to 800\,Myr.
We witness the longest time for the M2 model, which has the least eccentric orbit for merging galaxies and resulted in a relatively lower stellar density of its merger remnant.
On the other hand, we observe the shortest coalescence time for model M3, where the secondary-to-primary SMBH ratio is $q_\bullet=0.06$.
We also notice that, in general, SMBH binaries spend a significant span of their lifetime in the three-body scattering phase, which is in contradiction to our earlier study of overmassive SMBH mergers \citep{Khan+15}.
\citet{Khan+15} reported that SMBH coalescence times are very short once an SMBH binary forms following a galaxy merger, almost bypassing the three-body hardening phase.
In that study, the SMBH mass ratio was very small ($q_\bullet=0.008$), resulting in very high eccentricities that considerably shorten the SMBH binary lifetime.
In fact, we witness a similar trend of shortening the binary merger time of an SMBH for the case of a smaller mass ratio ($q_\bullet=0.06$) in run M3.

The initial setup of our galaxy merger simulations corresponds to an advanced stage of galaxy merging in a cosmological context.
To test the impact of a larger initial orbit, we ran a model in which the two merging galaxies start at an apocenter separation of 5\,kpc and a pericenter of 500\,pc.
This setup is identical to model M1, except for the larger orbital separation.
We find that the galaxy merger is delayed by approximately 75\,Myr; however, once the galaxies merge, the subsequent SMBH dynamics and merger times are practically indistinguishable from those in model~M1.
This outcome is expected, as the compact nature of the galaxies means that they experience significant tidal interactions primarily during the late stages of orbital evolution, when the pericenter passage becomes comparable to their effective radii.
We also find that changes in the half-mass radius and, consequently, in the central surface density closely match those observed in model~M1.

\subsection{SMBH Mergers vs.\ Galaxy Mergers}

Our simulations reveal that when it comes to the mechanism governing the transformation of LRDs into less dense ancestors of modern-day ellipticals, it is the SMBHs that are paramount, not the galaxies.
The overarching result of our five merger runs is that a nearly-equal SMBH mass ratio is required to efficiently scour out the cores of dense LRDs.
In contrast, the particular properties of the merging galaxies or the specifics of their orbital parameters are far less consequential in transforming LRDs.
This imbalance of importance between galaxies and SMBHs may help explain why we do not observe extant LRDs in the local Universe.

If the opposite were true (i.e., galaxy properties being more important than the SMBHs), the elimination of LRDs might not have been as complete.
In the early Universe ($z\sim7$), the slope of the galaxy stellar mass function was steep, i.e., less massive galaxies were more abundant than more massive galaxies \citep[see][Fig.~6]{Grazian:2015}.
Therefore, if a high secondary-to-primary galaxy mass ratio was needed to transform LRDs, then we might expect that LRDs would survive most encounters.
Although, new results from the \textit{JWST} are beginning to find that LRDs may live in a flatter high-mass part of the galaxy stellar mass function \citep[see][Fig.~6]{Weibel:2024}.
Nonetheless, we postulate that as long as the merger of an LRD with another galaxy results in the major merger of SMBHs, the density of the central region of the LRD will be significantly depleted, and the material will be driven toward the outer regions of the galaxy, increasing half-mass radius, and effectively ending its status as an LRD.

\subsection{Implications for Gravitational-wave Astronomy}

The inspiral of binary $10^8$--$10^9$\,$\mathrm{M}_\sun$ black holes featured here emit squarely in the ultra-low-frequency regime probed by pulsar timing arrays, although they would be too distant to individually resolve \citep{EPTA,PPTA,NANOGrav}.
In fact, these sources may have contributed to the recent pulsar timing detection of a louder-than-expected gravitational wave background \citep{NANOGrav:2023}.
The actual mergers of these black hole binaries would occur in the frequency window between pulsar timing and space-based gravitational wave detectors, such as the Laser Interferometer Space Antenna (LISA), an ESA/NASA mission set to launch in 2035~\citep{LISA:2023}.
If the black hole masses were $\lesssim$$10^7$\,$\mathrm{M}_\sun$, as in model 55604B, LISA would be able to observe the merger with signal-to-noise ratios in the hundreds, which would enable measurement of properties such as mass ratio, spin, and spin orientation, which are thought to constrain black seed formation and early growth scenarios \citep{redbook}.
Indeed, an early census has identified an overabundance of lower-mass black holes beyond $z>4$ \citep[e.g.][]{Maiolino24,Taylor24,Li:2025}.

\subsection{Gas Dynamics}\label{sec:gas_dynamics}

Our study does not include the effects of gas dynamics, we briefly discuss how this might influence our findings.
Gas-rich high-redshift galaxies can form compact stellar clusters either through gas instabilities \citep{Mayer2025} or via feedback-free starbursts \citep{Dekel2025}.
These clusters can subsequently inspiral toward the galactic center due to dynamical friction and dissolve through tidal interactions, thereby enhancing the central stellar density.
Such mechanisms are also conducive to the efficient formation and growth of SMBHs in dense stellar and gaseous environments.
However, it is reasonable to assume that our progenitor LRDs have already undergone this phase of intense star formation and SMBH growth, as suggested by their high stellar and black hole masses.

The spectral energy distributions of LRDs extend over several hundred million years, suggesting sustained star formation required to build the evolved stellar populations that give rise to the observed Balmer breaks \citep{Wang:2024}.
Furthermore, it has been suggested that the large stellar masses of LRDs can only be reconciled if a substantial fraction of baryonic matter has already collapsed as a result of an initially very high star formation rate \citep{Boylan2023, Boylan2025}.
The presence of gas in the central region can suppress the amplitude of the SMBH recoil oscillation and shorten the timescale to settle in the center \citep{Guedes2011}.
For our case, we have chosen a moderate value of recoil kick and we notice that the effective radius changes abruptly as the SMBH oscillates in the merger remnant and deposits energy to the surrounding medium.

\section{Conclusions}\label{sec:conclusions}

Our findings highlight the pivotal role that SMBH dynamics may play in resolving the tension between the extreme central densities of LRDs and the properties of low-redshift elliptical galaxies.
The preponderance of massive black holes in the LRD epoch may point to their necessity in early galaxy assembly. 
Our simulations prove that while galaxies assemble through major mergers, the SMBHs coalesce by ablating the stellar core, and as GW recoil kicks the newly merged SMBH out of the core, the central density drops even more.
By the end of our simulations, the effective radii, central surface, and 3D densities of the merger remnants reach values that are in line with galaxy observations at intermediate-to-low redshifts. 
These processes work on the right spatial and temporal scales to transform the centers of these extreme objects into more plausible low-redshift descendants.

\begin{acknowledgments}
This material is based on work supported by Tamkeen under the NYU Abu Dhabi Research Institute grant CASS.
This research has used NASA's Astrophysics Data System.
This research was carried out on the high-performance computing resources at New York University Abu Dhabi.
\end{acknowledgments}

\software{
\href{https://github.com/berczik/phi-GPU-mole}{\textcolor{linkcolor}{\texttt{$\phi$-GPU}}} \citep{Berczik:2011,Berczik:2013},
\href{https://github.com/matplotlib/matplotlib}{\textcolor{linkcolor}{\texttt{Matplotlib}}} \citep{matplotlib},
\href{https://github.com/numpy/numpy}{\textcolor{linkcolor}{\texttt{NumPy}}} \citep{numpy},
\href{https://pandas.pydata.org/}{\textcolor{linkcolor}{\texttt{Pandas}}} \citep{pandas},
\href{https://www.python.org/}{\textcolor{linkcolor}{\texttt{Python}}} \citep{Python}
}

\section*{ORCID iDs}

\begin{CJK*}{UTF8}{gbsn}
\begin{flushleft}
Fazeel Mahmood Khan \scalerel*{\includegraphics{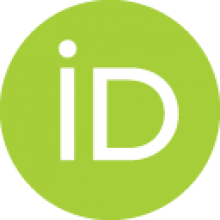}}{B}\\\url{https://orcid.org/0000-0002-5707-4268}\\
Benjamin L.\ Davis \scalerel*{\includegraphics{orcid-ID.png}}{B} \url{https://orcid.org/0000-0002-4306-5950}\\
Andrea Valerio Macci\`{o} \scalerel*{\includegraphics{orcid-ID.png}}{B} \url{https://orcid.org/0000-0002-8171-6507}\\
Kelly Holley-Bockelmann \scalerel*{\includegraphics{orcid-ID.png}}{B} \url{https://orcid.org/0000-0003-2227-1322}
\end{flushleft}
\end{CJK*}

\bibliography{bibliography}

\end{document}